\documentstyle[12pt]{article}

\renewcommand{\theequation}{{\arabic{section}.\arabic{equation}}}

\def\thesection{\arabic{section}}
\def\thesubsection{\arabic{section}.\arabic{subsection}.}
\def\appendices{\par 
\setcounter{section}{1} \setcounter{subsection}{0}
\def\thesection{\Alph{section}.}
\def\thesubsection{\thesection\arabic{subsection}}
\def\theequation{\Alph{section}.\arabic{equation}}
\nopagebreak
}
\catcode `\@=11
\def\@cite#1#2{#1\if@tempswa , #2\fi} 
\catcode `\@=12
\newcommand{\upcite}[1]{${}^{\mbox{\scriptsize (\cite{#1})}}$}
\newcommand{\onlinecite}[1]{$\cite{#1}$}

\begin{document}
\begin{center}
{\large Low-Lying Excited States of Quantum Antiferromagnets\\ 
on a Triangular Lattice}\footnote{To be published in J. Stat. Phys. {\bf 75} 
(1994) }
\bigskip

{\large Tsutomu Momoi}
\bigskip

Department of Physics, University of Tokyo, 
Hongo 7-3-1, Bunkyo-ku, Tokyo 113, Japan
\end{center}

\begin{abstract}
We study low-lying states of the {\it XY} and Heisenberg antiferromagnets 
on a triangular 
lattice  to clarify whether spontaneous symmetry breaking occurs at $T=0$ in 
the thermodynamic limit. Approximate forms of low-lying states are proposed, in 
which degrees of freedom of the sublattice magnetization and of the chirality 
are separated. It is shown that low-lying states can be accurately 
described with the present approximation. It was argued that low-lying states 
play an important role in symmetry breaking. 
With help of this approximation, we discuss the contribution of low-lying 
states to symmetry breaking of two types, namely creation of the spontaneous 
sublattice magnetization and the spontaneous chirality. 
Furthermore, to show an evidence for the occurrence of the symmetry breaking, 
we numerically study the low-lying states of finite systems of the 
{\it XY} and Heisenberg antiferromagnets. It is found that 
the necessary conditions for the symmetry breaking to occur are 
satisfied in these models.
\end{abstract}
\bigskip


\setcounter{equation}{0}
\section{Introduction }
Many authors have studied ground-state properties of quantum
antiferromagnets, e.g., the Heisenberg and {\it XXZ} models, 
on a triangular lattice. The Hamiltonian of the {\it XXZ} antiferromagnet is
given by 
\begin{equation}
H=J\sum_{\langle i,j \rangle}(S_i^xS_j^x+S_i^yS_j^y+\Delta S_i^zS_j^z)
\end{equation}
where the summation runs over all the nearest-neighbour pairs. 
Most of the studies are based on one of the following two viewpoints. 
The first one is that the system
is disordered. Anderson proposed the RVB state\upcite{Anderson,Fazekas}, 
and Kalmeyer and Laughlin discussed a spin-liquid state\upcite{Kalmeyer} 
from this
aspect. The second one is that the system has a long-range order with 
the so-called $120^\circ$ structure, 
though quantum fluctuations reduce the sublattice magnetization. 
Trial wave functions which
have the $120^\circ$ structure are discussed by 
Miyashita\upcite{Miyashita1984}, and by Huse and Elser\upcite{HuseE}. 
The spin-wave expansions\upcite{Oguchi,Miyake,Momoi,Leung}
and the series expansion\upcite{Singh} suggested the existence of a long-range
order. The present author and Suzuki\upcite{Momoi} presented 
the exact ground states of the {\it XXZ} model with $\Delta=-0.5$, 
most of which have the $120^\circ$ structures. We also studied the
{\it XXZ} antiferromagnet in the region $-0.5<\Delta\le 1$ using the 
spin-wave expansion and showed that the quantum fluctuations are enhanced, as
$\Delta$ becomes large. The quantum effect is strong in the Heisenberg model.
The estimates of the ground-state energy in the studies based on these 
pictures are almost equal to each other in the Heisenberg model. Thus it is
difficult to judge which picture is correct. 

Apart from the above arguments, several 
authors\upcite{Nishimori1988,Bernu,Leung} 
studied ground states in finite systems using the exact-diagonalization method. 
There are two papers\upcite{Bernu,Leung} which reported the calculations up to 
$N=36$. Conclusions do not coincide with each other. 
Bernu, Lhuillier and Pierre\upcite{Bernu} reported 
that the sublattice order is almost equal to that obtained by the spin-wave 
expansion. 
Leung and Runge's result\upcite{Leung} is, however, 
contradictory to the existence of the sublattice order.
These two groups fit the data in different ways. 

Understanding of symmetry breaking has progressed by studies of low-lying
excited states. In quantum antiferromagnets the ground state of a finite 
system is, in many cases, unique 
and symmetric. In the thermodynamic limit, symmetry is broken by
making a linear combination of the ground state and low-lying excited states. 
Horsch and von der Linden\upcite{Horsch} first showed that, if the Heisenberg
antiferromagnet has a N\'eel order, there exists a 
low-lying excited state whose energy converges to that of the ground state. 
Koma and Tasaki\upcite{KomaB} rigorously
proved that, if the system has a long-range order, low-lying states of growing 
numbers converge to the ground state in the thermodynamic limit.
Several numerical calculations showed the existence of these low-lying excited
states in the spin $1/2$ Heisenberg antiferromagnet on a square 
lattice\upcite{Tang} and on a triangular lattice\upcite{Bernu}.

Bernu {\it et al.}\upcite{Bernu} determined the quantum numbers of low-lying
states which construct the ordered infinite-volume ground states 
in the Heisenberg antiferromagnet on the triangular lattice. 
They studied finite systems and found the existence of a whole set of 
low-lying states.  Azaria {\it et al.}\upcite{Azaria1993} studied 
the finite-size dependence of these
low-lying states. They found that these states satisfy the scaling property 
which indicates the existence of a long-range order. 

In this paper we consider the {\it XY} and Heisenberg antiferromagnets 
on the triangular lattice. We study low-lying states to clarify 
whether spontaneous symmetry-breaking occurs at $T=0$ in the thermodynamic 
limit. We give approximate forms of low-lying states, in which 
the degrees of freedom of the sublattice magnetization and the chirality 
are separated. We show that the low-lying states can be accurately 
described with the present approximation. With help of our
approximation we discuss how 
rearrangements of the low-lying states bring on symmetry breaking. 
In quantum antiferromagnets 
on the triangular lattice, there are two types of symmetry breaking; 
In the Heisenberg antiferromagnet both the sublattice magnetization and the 
chirality relate to the breaking of the 
$O$(3) symmetry and in the {\it XY} antiferromagnet 
the $O$(2)$\times Z_2$(chiral) symmetry can be broken. 
These symmetry breakings relate to each other. 
Using our approximation we can understand breakdown of each symmetry 
independently. 
Furthermore, studying the low-lying states of finite systems with the numerical
diagonalization method, 
we obtain an evidence of occurrence of the symmetry breaking. 
In the {\it XY} model the necessary conditions for the symmetry to break are 
fully satisfied. We conclude that there exists long-range order in the 
thermodynamic limit. 
In the Heisenberg model, since the system size studied here 
is too small, we can not still obtain a definite conclusion 
about the existence of long-range order, but there are many evidences which
support the existence of the long-range order, as discussed in refs. 
\onlinecite{Bernu} and \onlinecite{Azaria1993}.

The relation between the long-range order parameter $\sigma$ and
the spontaneous sublattice magnetization $m$ is also argued, though this is not
the main part of the present paper. It has been discussed\upcite{KomaA} that, 
for bipartite systems, the relation is given by 
$m=\sqrt3 \sigma$ in the Heisenberg antiferromagnet and by $m=\sqrt2\sigma$
in the {\it XY} antiferromagnet. 
In the antiferromagnets on the triangular lattice, however, the
factor of the relation becomes $\sqrt 2$ times as large as the above. 
We obtain $m=\sqrt6 \sigma$ for the Heisenberg model and $m=2\sigma$ for the 
{\it XY} model. 

Contents of this paper are the following. Approximate low-lying states are
formally given and properties of these states are discussed 
in Section~\ref{sec2}. Some results for the low-lying states of ferromagnets,
which are used in Section~\ref{sec2} to discuss the properties of the
approximation, are shown in Appendix A. The mechanism of symmetry breaking
is discussed in Section~\ref{sec:SB_mechanism}.
In Section~\ref{sec:evidence} we show an evidence of the symmetry breaking 
studying finite systems numerically. 
Section~\ref{sec:summary} contains summary and discussions. 
The relations between the symmetry breaking and the long-range order parameter 
are discussed in Appendix B. 

\setcounter{equation}{0}
\section{An approximation for low-lying states }\label{sec2}
We study properties of the low-lying states of the quantum 
antiferromagnets on a finite-volume triangular lattice, using approximate
states. In antiferromagnets on finite systems the ground state is symmetric. 
Symmetry breaking in an infinite-volume limit is discussed in the next section.

We consider the quantum {\it XXZ} antiferromagnets on the finite-volume 
lattice $\Lambda$ with a periodic boundary condition. The size of the
system is $N$, where $N=3m$ with an integer $m$. The Hamiltonian is given by 
\begin{equation}\label{Hamiltonian_XXZ}
H=J\sum_{\langle i,j \rangle \in \Lambda}
(S^x_i S^x_i + S^y_i S^y_i + \Delta S^z_i S^z_i ),
\end{equation}
where the summation runs over all the nearest-neighbor sites and 
the symbol $\Delta$ denotes the anisotropy of the $z$-component interactions. 
The models in the region $-0.5 < \Delta \le 1$ are considered.  

We concentrate on low-lying states which have sublattice-translational 
invariance and $C_{3V}$-invariance. (We denote the point group of
the $120^\circ$ rotation and reflection of the lattice as $C_{3V}$.)  
Bernu {\it et al.}\upcite{Bernu} reported that 
there are many low-lying states of this type. 
Here we classify the states according to eigenvalues of the $60^\circ$ 
rotation ($C_6$). 
We call the class of $C_6$-symmetric states as the type $\alpha$ and
the class of $C_6$-antisymmetric states as the type $\beta$. 

As an approximation for the lowest states of the type $\alpha$ and of the type
$\beta$ in the $S^z_{\rm total} =n$ (or $n+1/ 2$) subspace, 
we consider the following states: 
\begin{equation}
\label{st_approx}
\vert n\alpha\rangle 
= \frac{(U+U^\dagger)\vert {\rm F}n\rangle}
{\Vert(U+U^\dagger)\vert{\rm F}n\rangle\Vert},
\mbox{\hspace*{1cm}}
\vert n\beta\rangle 
= \frac{(U-U^\dagger)\vert {\rm F}n\rangle}
{\Vert(U-U^\dagger)\vert {\rm F}n\rangle\Vert}
\end{equation}
for $n=0,\pm 1,\pm 2,\ldots$, where the symbol $U$ denotes the unitary operator 
\begin{equation}
U=\exp\Bigl(i{2\pi\over3}\sum_{i\in B} S_i^z-i{2\pi\over3}\sum_{i\in C} S_i^z
\Bigr) 
\end{equation}
and $\Vert (U \pm U^\dagger ) |{\rm F}n \rangle \Vert = 
\langle {\rm F}n |(U \pm U^\dagger )^2 |{\rm F}n \rangle^{1/2}$. 
The state $\vert {\rm F}n \rangle$ denotes the lowest state in the 
$S^z_{\rm total}=n$ (or $n+1/2)$ subspace of the ``ferromagnetic'' 
{\it XXZ} model on $\Lambda$ which is defined by the Hamiltonian
\begin{equation}\label{Hamiltonian_appro}
H_{\rm F}=-J\sum_{\langle i,j \rangle \in \Lambda}
(S_i^xS_j^x+S_i^yS_j^y-2\Delta S_i^zS_j^z). 
\end{equation}
Here the parameter $\Delta$ is set equal to that in (\ref{Hamiltonian_XXZ}). 
We denote the eigenvalue of $H_F$ for the state $\vert {\rm F}n\rangle$ as 
$E_{{\rm f}n}$. 

We first discuss the properties of the above approximate states 
(\ref{st_approx}). 
After that we verify that this approximation is good. 

Here we discuss the meaning of the referred model (\ref{Hamiltonian_appro}). 
The states 
$\vert n\alpha \rangle$ and $\vert n\beta \rangle$ are linear combinations
of $U\vert {\rm F}n \rangle$ and $U^\dagger\vert {\rm F}n \rangle$. 
The unitary operator $U$ transforms the Hamiltonian (\ref{Hamiltonian_XXZ})
into the form 
\begin{equation}\label{Hamiltonian_UHU}
UHU^\dagger=-{1\over 2}J\sum_{\langle i,j \rangle}
(S_i^xS_j^x+S_i^yS_j^y-2\Delta S_i^zS_j^z)
-{\sqrt3\over2}J\sum_{\langle i\to j \rangle}(S_i^xS_j^y-S_i^yS_j^x).
\end{equation}
Here the symbol $i\to j$ goes from the sublattice A to B, B to C, and C to A.
The first term in (\ref{Hamiltonian_UHU}) gives the
Hamiltonian (\ref{Hamiltonian_appro}). 
Thus the present approximation corresponds to neglecting the second term in 
(\ref{Hamiltonian_UHU}). 

In the thermodynamic limit of the ferromagnet (\ref{Hamiltonian_appro}), 
only $O$(2) symmetry breaking can take place, since it has no frustration. 
The classical limit of the ferromagnet
(\ref{Hamiltonian_appro}) in the region $-0.5< \Delta \le 1$ has
ferromagnetically ordered ground states, in which all spins are lying in the
$XY$ plane\upcite{Miyashita1986}. The spin $1/2$ ferromagnetic $XY$ model
($\Delta=0$) on the triangular lattice has been studied by using the
exact-diagonalization method\upcite{Nishimori1988} and the spin-wave
theory\upcite{Nishimori1985,Weihong}.
The results indicate the existence of the ferromagnetic long-range order. 
As $\Delta$ increases in the ferromagnet (\ref{Hamiltonian_appro}), quantum
fluctuations are enhanced and the magnetization is reduced. 

As shown in Appendix A, several exact results can be obtained for the
ferromagnet (\ref{Hamiltonian_appro}) on the finite-volume lattice $\Lambda$. 
The ground state is unique and it exists in the 
$S^z_{\rm total}= 0$ subspace. Then it has $O$(2)-rotational invariance. 
(When $N$ is an odd number, there exists trivial degeneracy; the states 
$| {\rm F}0 \rangle$ and $|{\rm F}- \;1 \rangle$ are degenerate.) 
The lowest state in the $S^z_{\rm total}=n$ subspace, which we denote as $|{\rm
F}n\rangle$, is unique. The coefficients of $|{\rm F}n\rangle$ in the basis
which are eigenstates of $\{$~$S^z_i$~$\}$ are nonnegative. 
All states $\{$~$|{\rm F} n \rangle$~$\}$ 
are translationally invariant and $C_{6V}$-invariant. (See Appendix A.)

Using these rigorous results for the low-lying states of the ferromagnet, we can
determine the spatial quantum numbers of the approximate states. We have
verified in the small clusters that the spatial quantum numbers of the
approximate states (\ref{st_approx}) are the same as those of the true states. 
Bernu {\it et al.}\upcite{Bernu} investigated the spatial quantum numbers of the 
low-lying states for the Heisenberg antiferromagnet and classified the 
low-lying states into three classes; 
$\Gamma_1$(states with $k=0$ and even under inversion), 
$\Gamma_2$(states with $k=0$ and odd under inversion) and 
$\Gamma_3$(states with $k=\pm (4\pi/3,0)$). 
When $S^z_{\rm total}(=n)=3l$ or $S^z_{\rm total}(=n+1/2)=3l+3/ 2$ with 
an integer $l$, $|n \alpha \rangle$ and $|n \beta \rangle$ 
belong to $\Gamma_1$ and $\Gamma_2$, respectively. 
Otherwise, both $|n \alpha \rangle$ and $|n \beta \rangle$ belong to 
$\Gamma_3$. 
As we show in the next section, our classification is useful 
in discussing the symmetry breaking about the sublattice magnetization 
and the chirality separately. 

We now discuss accuracy of our approximation. We compare the approximate 
low-lying states with the true ones, numerically diagonalizing finite systems. 
First we note that, as we have already mentioned above, the spatial 
quantum numbers of our approximate low-lying states are
equal to those of the true states: 
They are invariant under any sublattice
translation and any translation of $C_{3V}$, and they have the same momentum as
the true states. 
In the true system there exist low-lying states of two types corresponding to 
the types $\alpha$ and $\beta$. 
Secondly the approximate low-lying states have accurate
expectation values of the energy, which are at most $2\%$ higher than the exact
values in the {\it XY} model and at most $7\%$ higher in the Heisenberg model. 
Finally we calculated overlaps between the exact low-lying states and the
approximate ones. The results are shown in Table~\ref{Table:overlap}. 
In the {\it XY} model the approximate states have more than
$90\%$ overlaps with the exact states and in the Heisenberg model more than
$70\%$. These situations are the same for the low-lying
states in the $S^z_{\rm total}=1$ and $S^z_{\rm total}=2$ subspaces. 
Thus we find that the exact low-lying spectrum has the two-fold structures, 
as shown in Fig.~1, and that the true low-lying states can be accurately 
described with the present approximation. 
\begin{table}[bt]
\caption{Overlaps between the ground state $|0\rangle$ and the 
approximate state $|0\alpha\rangle$, and between the first excited 
state $|1\rangle$ and the approximate one $|0\beta\rangle$.}

\label{Table:overlap}
\newcommand{\lw}[1]{\smash{\lower2.ex\hbox{#1}}}
\begin{center}
\begin{tabular}{cl|cccc}
\hline
\hline
\multicolumn{2}{c|}{Size {\it N}}   & 3   & 9      &   12   &   21   \\
\hline
\lw{\it XY} & $|\langle0|0\alpha\rangle|^2$ & 1.0 & 0.9797 & 0.8472 & 0.9181 \\
            & $|\langle1| 0\beta\rangle|^2$ & 1.0 & 0.9797 & 0.9988 & 0.9181 \\
\hline
\lw{Heisenberg}&$|\langle0|0\alpha\rangle|^2$& 1.0 & 0.7881 & 0.7997 & 0.4643\\
               &$|\langle1|0\beta\rangle|^2$ & 1.0 & 0.7881 & 0.9871 & 0.4643\\
\hline
\hline
\end{tabular}
\end{center}
\end{table}

The approximate ground state, $\vert 0\alpha \rangle$, is 
a generalization of the trial state of Betts and Miyashita\upcite{Betts} for
the {\it XY} antiferromagnet. In the case of $\Delta=-0.5$, 
$\vert 0\alpha \rangle$ and $\vert 0\beta \rangle$ belong to the exact ground 
states, which were presented by the author and Suzuki\upcite{Momoi}. 
This fact partially gives the reason for accuracy of the state 
$U\vert {\rm F}0\rangle$. 

An advantage of our approximation is that the degrees of freedom of 
the $Z_{2}$(chiral) symmetry and the $O$(2) symmetry are separated in it. 
Then we can discuss breakdown of each symmetry independently. 
The parts of the unitary operator, namely $(U\pm U^\dagger)$, can describe the 
chiral symmetry breaking and the parts of the low-lying states of the
ferromagnet, namely $\vert {\rm F}n \rangle$ 
($n=0,\pm 1,\pm 2,\ldots)$, can display the breakdown of the $O$(2) symmetry. 

Using this approximation, we can estimate the expectation values of the energy. 
The ground-state energy is calculated as 
\begin{equation}
\langle 0\alpha\vert H \vert0\alpha \rangle
={ E_{{\rm f}0} + 4E_{{\rm f}0}\langle {\rm F}0\vert U\vert{\rm F}0 \rangle 
-6\Delta J\langle {\rm F}0\vert 
U(\sum_{\langle i,j \rangle}S^z_i S^z_j )\vert{\rm F}0 \rangle  
\over 2+2\langle {\rm F}0\vert U\vert{\rm F}0 \rangle},
\end{equation}
where we have used the relations $UHU^\dagger + U^\dagger HU=H_{\rm F}$ and 
$UHU=H_{\rm F}U^\dagger + U^\dagger H_{\rm F} 
-3\Delta JU^\dagger\sum_{\langle i,j \rangle}S_i^z S_j^z$. 
The terms $\langle {\rm F}0\vert U\vert{\rm F}0 \rangle$ and 
$\langle {\rm F}0\vert U(\sum_{\langle i,j \rangle}S^z_i S^z_j) 
\vert{\rm F}0 \rangle$ correspond to the transition probabilities 
from the ferromagnetic ground state, $\vert{\rm F}0\rangle$, 
to the antiferromagnetic state, $U\vert {\rm F}0 \rangle$. 
In the thermodynamic limit, these values are
vanishing. Thus the ground-state energy is estimated as 
\begin{equation}
\langle 0\alpha\vert H \vert0\alpha \rangle
\simeq{E_{{\rm f}0} \over 2}.
\end{equation}
Next we discuss the energy gaps. The energy gap between the ground state 
(the type $\alpha$) and the lowest state of the type $\beta$ is almost 
vanishing as
\begin{eqnarray}
\langle 0\beta\vert H\vert0\beta \rangle
-\langle 0\alpha\vert H\vert0\alpha \rangle
&=&{-3E_{{\rm f}0}\langle {\rm F}0\vert U\vert{\rm F}0 \rangle 
+6\Delta J\langle {\rm F}0\vert 
U(\sum_{\langle i,j \rangle}S^z_i S^z_j )\vert{\rm F}0 \rangle  
\over 1-\langle {\rm F}0\vert U\vert{\rm F}0 \rangle^2}\nonumber\\
&\simeq&0.
\end{eqnarray}
The energy gap between the ground state of $S^z_{\rm total}=0$ and the 
lowest state of $S^z_{total}=1$ is estimated as 
\begin{equation}
\langle 1\alpha\vert H\vert1\alpha \rangle
-\langle 0\alpha\vert H\vert0\alpha \rangle
\simeq{1\over2}(E_{{\rm f}1} - E_{{\rm f}0}).
\end{equation}
Relations between these energy gaps and the symmetry breaking of two
types, namely creation of the spontaneous chirality and of the spontaneous
sublattice magnetization, are discussed in Section~\ref{sec:SB_mechanism}. 

We give a remark for the case of the Heisenberg antiferromagnet. 
Since the model is isotropic, 
a slight modification is necessary for the approximate ground state, 
$\vert 0\alpha_{(\Delta=1)} \rangle$. 
We only give the form of an approximate ground state as follows: 
\begin{equation}\label{st_approxH}
\vert 0\alpha^\prime \rangle = {1 \over 4\pi}
\int d\Omega R(\mbox{\boldmath $\Omega$})
\vert 0 \alpha_{(\Delta =1)}\rangle
\Bigl/\Bigl\Vert{1 \over 4\pi}
\int d\Omega R(\mbox{\boldmath $\Omega$})
\vert 0 \alpha_{(\Delta =1)}\rangle\Bigr\Vert, 
\end{equation}
where {\boldmath $\Omega$} is the unit vector with the spherical coordinates 
$(\theta,\varphi)$ and 
\begin{equation}
R(\mbox{\boldmath $\Omega$})=\exp \Bigl(i\varphi\sum_i S_i^z\Bigr)
\exp \Bigl(i\theta\sum_j S_j^y\Bigr).
\end{equation}
We used the state $\vert 0\alpha_{(\Delta=1)} \rangle$ 
to compare our approximation with the true ground state of 
the Heisenberg model, for it is not easy to use $\vert 0\alpha^\prime \rangle$ 
in numerical calculations. 

\setcounter{equation}{0}
\section{Contribution of the low-lying states to the symmetry breaking }
\label{sec:SB_mechanism}
Several authors\upcite{Bernu,Horsch,KomaB,Tang} have discussed the mechanism of 
symmetry breaking in quantum Heisenberg antiferromagnets. 
It has been argued that, as the system size is enlarged, many 
low-lying states converge to the ground states and that linear combinations of 
these states make the symmetry breaking occur. 
In this section we study the roles played by the low-lying states of the two
types in the symmetry breaking. 
Here we assume that the ground states of 
an infinite system have a long-range order and that the symmetry breaks down. 
In Section~\ref{sec:evidence} we show an evidence of occurrence of the symmetry 
breaking. 

In antiferromagnets on a triangular lattice, there are two types of symmetry
breaking. Then the mechanism of symmetry breaking is complex. 
In the {\it XY} model, or in the {\it XXZ}
model for $-0.5 < \Delta < 1$, the $O$(2) symmetry and the 
$Z_2$(chiral) symmetry can break independently. 
In the Heisenberg model, both the spontaneous sublattice 
magnetization and the spontaneous chirality correspond to the breakdown of the
$O$(3) symmetry. 

As shown in Section~\ref{sec2}, the low-lying states on a finite-volume
triangular lattice have the two-fold structures. 
The purpose of this section is to clarify the relations between the symmetry
breaking of the two types and the low-lying states of the types $\alpha$ and
$\beta$. Our approximation (\ref{st_approx}) helps us to understand the 
symmetry breaking of the two types independently, since the
degrees of freedom of the sublattice magnetization and of the chirality are
separated in it. 

In an infinite system we can define four types of ground states, which are 
classified by existence of the spontaneous sublattice magnetization or the 
spontaneous chiral order. Here we show the definitions of the four types. \\
(1) {\it Symmetric states} ({\it Mixed states}),
\begin{equation}\label{def:mixed_state}
\langle \cdots \rangle_1
=\lim_{N \uparrow \infty}\lim_{\beta \uparrow \infty}
\frac{\mbox{Tr}[\cdots \exp(-\beta H)]}
{\mbox{Tr}[\exp(-\beta H)]}=\omega (\cdots).
\end{equation}
(2) {\it States in which only the spontaneous chirality exists}, 
\begin{equation}
\langle \cdots \rangle_2
=\lim_{B \downarrow 0}\lim_{N \uparrow \infty}\lim_{\beta \uparrow \infty}
\frac{\mbox{Tr}[\cdots \exp(-\beta (H-BQ^z))]}
{\mbox{Tr}[\exp(-\beta (H-BQ^z))]}
\end{equation}
where $Q^z$ denotes the $z$-component of the chirality order-parameter operator 
\begin{equation}
\mbox{\boldmath $Q$}=\frac{2}{\sqrt3}\sum_{\langle i \rightarrow j \rangle} 
(\mbox{\boldmath $S$}_i\times \mbox{\boldmath $S$}_j).
\end{equation}
The summation is defined in the same way as in eq.~(\ref{Hamiltonian_UHU}).\\
(3) {\it States in which only the spontaneous magnetization exists}, 
\begin{equation}\label{def:smag_state}
\langle \cdots \rangle_3
=\lim_{B \downarrow 0}\lim_{N \uparrow \infty}\lim_{\beta \uparrow \infty}
\frac{\mbox{Tr}[\cdots 
\exp\{-\beta (H-B(U\sum_i S_i^x U^\dagger + U^\dagger \sum_i S_i^x U))\}]}
{\mbox{Tr}[\exp\{-\beta (H-B(U\sum_i S_i^x U^\dagger + U^\dagger\sum_i S_i^x U
))\}]}.
\end{equation}
(4) {\it Pure states}, 
\begin{equation}\label{def:pure_state}
\langle \cdots \rangle_4
=\lim_{B \downarrow 0}\lim_{N \uparrow \infty}\lim_{\beta \uparrow \infty}
\frac{\mbox{Tr}[\cdots \exp(-\beta (H-BU^\dagger \sum_i S^x_i U))]}
{\mbox{Tr}[\exp(-\beta (H-BU^\dagger \sum_i S^x_i U))]}=\tilde{\omega}
(\cdots).
\end{equation}
Here the equilibrium states are defined as functionals of expectation 
values on an operator space. (This is familiar in studies of 
infinite-volume systems.) In the following, we show, using the approximate 
states (\ref{st_approx}), how each ground state is constructed of low-lying
states. 

{\sf (1) Symmetric states (Mixed states).} 
By taking the thermodynamic limit of the finite-volume ground state, 
we get the symmetric ground state (\ref{def:mixed_state}). 
No symmetry is broken in it. 

{\sf (2) States in which only the spontaneous chirality exists.} 
We obtain the state $\langle\cdots\rangle_2$ by taking the 
thermodynamic limit under an infinitesimal effective field which is conjugate
to the chiral order parameter. 

An approximation of this state is given by $|2\rangle =U|{\rm F}0 \rangle$. 
There remains the $O$(2) invariance in it. 
Both Miyashita's trial function\upcite{Miyashita1984} and 
the variational function by Huse and Elser\upcite{HuseE} belong to this type of 
ground states. 

In the $XY$-like model, the state $U\vert \mbox{F}0 \rangle$ is 
constructed by making a linear combination of 
$\vert 0\alpha\rangle$ and $\vert 0\beta\rangle$, 
\begin{equation}
\{\vert 0\alpha \rangle,\;\; 
\vert 0\beta \rangle\} {\hbox to 1cm{\rightarrowfill}} 
\{U\vert \mbox{F}0 \rangle,\;\;
U^\dagger\vert \mbox{F}0 \rangle\}.
\end{equation}
Thus with help of our approximation we find that, if the ground state 
(the type $\alpha$) and the lowest state of the type $\beta$ become degenerate 
in the infinite-volume limit and if they are rearranged between themselves, 
the $Z_2$ symmetry of the chirality is broken. 

In the Heisenberg model, the degree of freedom of the chirality is continuous
and the ground state is isotropic, as mentioned in (\ref{st_approxH}). 
Then two states are not enough to break entirely the symmetry of the chirality. 
Koma and Tasaki\upcite{KomaB} proposed an approximate state
which describes breakdown of continuous symmetry and they argued that the
thermodynamic limit of this approximate state is a pure state, in which
continuous symmetry is broken. 
By applying their argument to the present case, an approximate state 
which has fully-ordered spontaneous chirality can be constructed in the form 
\begin{equation}
{1\over \sqrt{2k+1}}\left\{ \vert \Phi_{\rm GS} \rangle 
+ \sum_{n=1}^{k} \left( \frac{(Q^+)^n \vert \Phi_{\rm GS} \rangle} 
{\Vert (Q^+)^n \vert \Phi_{\rm GS} \rangle\Vert} 
+ \frac{(Q^-)^n \vert \Phi_{\rm GS} \rangle}
{\Vert (Q^-)^n \vert \Phi_{\rm GS} \rangle\Vert}\right) \right\}. 
\label{state_appro2H}
\end{equation}
where $Q^\pm=Q^x\pm iQ^y$ and $| \Phi_{\rm GS} \rangle$ denotes the ground
state. The number $k$ is of $o(N)$. 
This state has a nonvanishing expectation value of
the operator $Q^x$. By rotating all the spins through the angle $\pi/2$
about the $y$ axis, 
we obtain an approximation for the state $\langle\cdots\rangle_2$. 
Following the arguments by Koma and Tasaki\upcite{KomaB}, 
the infinite-volume limit of this 
approximate state become $\langle \cdots \rangle_2$. 
The states $(Q^\pm )^{2m} \vert \Phi_{\rm GS}\rangle$ 
belong to the type $\alpha$ and $(Q^\pm )^{2m+1} \vert \Phi_{\rm GS}\rangle$ to
$\beta$. Thus the state $\langle \cdots \rangle_2$ is constructed by a linear 
combination of low-lying states of the type $\alpha$ and of the type $\beta$. 

In our approximation the expectation value of the spontaneous chirality is 
calculated as
\begin{eqnarray}
\langle 2 \vert Q^z \vert 2 \rangle
&=& {E_{{\rm f}0} \over J}
+{E_{{\rm f}0} \over 3J }\langle \mbox{F}0 \vert U \vert \mbox{F}0 \rangle 
-2\Delta\langle \mbox{F}0 \vert \sum_{\langle i,j \rangle}S_i^z S_j^z 
\vert \mbox{F}0 \rangle \nonumber\\
&\simeq& {E_{{\rm f}0} \over J}
-2\Delta\langle \mbox{F}0 \vert 
\sum_{\langle i,j \rangle}S_i^z S_j^z \vert \mbox{F}0 \rangle 
\label{ev:SB_ch}
\end{eqnarray}
where the relations 
${\displaystyle UHU^\dagger={1\over2}H_{\rm F}-{3\over4}JQ}$ and 
$H=-H_{\rm F}+3J\Delta \sum_{\langle i,j \rangle}S_i^zS_j^z$ are used and 
$E_{{\rm f}0}$ denotes the ground state energy of the ferromagnet. 
As $\Delta$ is increased from $-0.5$, 
the first term in (\ref{ev:SB_ch}) becomes large by quantum effects. 
The second term in (\ref{ev:SB_ch}), 
$\langle \mbox{F}0 \vert \sum_{\langle i,j \rangle}S_i^z S_j^z 
\vert \mbox{F}0 \rangle$, behaves as follows: At $\Delta=-0.5$
this term is vanishing, since 
there is no correlation between up and down spins in the ground state $\vert
{\rm F}0 \rangle$. As the
parameter $\Delta$ increases, nearest-neighbor pairs of up and down spins are 
favored. Then the term becomes large. Thus at $\Delta=0$ the chirality is
enhanced by the quantum effect of the first term and near $\Delta=1$ the 
value is reduced by contribution of the second term. 

{\sf (3) States in which only the spontaneous magnetization exists.} 
In the state $\langle \cdots \rangle_3$ there exists the spontaneous 
magnetization, though there exists no spontaneous chirality. 

In the {\it XY}-like model, the $O$(2)-symmetry breaking occurs in the
following way. First we consider the case of the ferromagnetic $XXZ$ 
model~(\ref{Hamiltonian_appro}). As shown in Appendix A, the finite-volume
ground state of the ferromagnet is unique and it has $O$(2)-rotational
invariance. It was shown in ref.~\onlinecite{KomaB} that the low-lying 
states $|{\rm F}n\rangle$ ($n=\pm 1,\pm 2, \dots ,\pm o(N))$ 
converge to the ground state and pure infinite-volume
ground states are constructed by taking linear combinations of the low-lying
states. We denote one of the ferromagnetically ordered ground states as 
$|\mbox{F-order}\rangle$, where 
$\langle\mbox{F-order}|S^x_i|\mbox{F-order}\rangle \ne 0$
and $\langle\mbox{F-order}|S^y_i|\mbox{F-order}\rangle=0$. 
In the case of the antiferromagnet, with help of the approximate low-lying
states, we can understand the mechanism of the $O$(2)-symmetry breaking on the 
analogy of the ferromagnet. Under the effective field $B$ in
(\ref{def:smag_state}), the low-lying states of the type $\alpha$ have an 
energy lower than those of $\beta$ have. 
An approximate state for $\langle\cdots\rangle_3$
is given by $(U+U^\dagger) |\mbox{F-order}\rangle$, which is constructed by 
making a linear combination of $(U+U^\dagger)|{\rm F} n \rangle$ $(n=0,\pm 1,\pm
2,\dots)$.
In the same way, by using low-lying states of the type $\beta$,
another state in which only the spontaneous magnetization exists is 
constructed as $(U-U^\dagger)\vert \mbox{F-order}\rangle$. 
Thus we find that, if the low-lying states of the same type become degenerate 
in the infinite-volume limit and if rearrangements occur between them, 
the breakdown of the $O$(2) symmetry occurs. 

For the Heisenberg model, by applying the approximation of Koma and
Tasaki\upcite{KomaB} to this model, an approximate state of 
$\langle \cdots \rangle_3$ is written as 
\begin{equation}
{1\over \sqrt{2k+1}}\left\{ \vert \Phi_{\rm GS} \rangle 
+ \sum_{n=1}^{k} \left( \frac{(O^+)^n \vert \Phi_{\rm GS} \rangle} 
{\Vert (O^+)^n \vert \Phi_{\rm GS} \rangle\Vert} 
+ \frac{(O^-)^n \vert \Phi_{\rm GS} \rangle}
{\Vert (O^-)^n \vert \Phi_{\rm GS} \rangle\Vert}\right) \right\} 
\end{equation}
where $O^+=U\sum_i S^+_i U^\dagger + U^\dagger \sum_i S^+_i U$. All the
low-lying states, $(O^+)^m \vert \Phi_{\rm GS} \rangle$ and 
$(O^-)^m \vert \Phi_{\rm GS} \rangle$, belong to the type
$\alpha$. Thus We can obtain the state $\langle \cdots \rangle_3$ by taking a
linear-combination of the low-lying states of the same type. 

{\sf (4) Pure states.} The pure infinite-volume ground states have the same 
structure as the ground states of the classical model. 
The $O$(2) symmetry and $Z_2$ symmetry are broken in the {\it XY} model, 
and the $O$(3) symmetry of the sublattice magnetization and that of the 
chirality are broken in the Heisenberg model. 

To construct the pure infinite-volume ground states, the symmetry breaking of 
two types discussed in the above should occur; 
A number of low-lying states of both types, $\alpha$ and $\beta$, 
become degenerate to the ground state and rearrangements occur between them. 
This is consistent with the previous arguments by Bernu 
{\it et al}.\upcite{Bernu}, and Koma and Tasaki\upcite{KomaB}. 

By discussing the symmetry breaking of two types separately, we find that
the spontaneous sublattice magnetization is created by rearrangements of the
low-lying states of the same type and that the spontaneous chirality is by
rearrangements of pairs of low-lying states of the type 
$\alpha$ and of $\beta$.

It is possible that, if long-range order of one type exists in the true system,
only one type of symmetry breaking occurs in the thermodynamic limit. 
Though the result from the renormalization group\upcite{Azaria1992} indicates 
existence of one critical
point (critical spin $S_{\rm c}$), our arguments suggest that symmetry breaking
can occur separately and that an intermediate phase can appear in which 
symmetry breaking of only one type occurs. 

An approximation for this state is given by 
\begin{equation}
\vert 4 \rangle=U \vert \mbox{F-order} \rangle
\end{equation}
and the expectation values are related with the quantities of the ferromagnet in
the forms 
\begin{equation}\label{eq_M_pure}
\langle 4 \vert U \sum_i S^x_i U^\dagger \vert 4 \rangle 
= \langle \mbox{F-order} \vert \sum_i S^x_i \vert \mbox{F-order} \rangle
\end{equation}
and
\begin{equation}
\langle 4 \vert Q^z \vert 4 \rangle=\langle 2 \vert Q^z \vert 2 \rangle.
\end{equation}

\setcounter{equation}{0}
\section{Evidence of symmetry breaking }\label{sec:evidence}
In this section we numerically study low-lying states of finite systems  
to verify the occurrence of 
symmetry breaking which was discussed in Section~\ref{sec:SB_mechanism}. 
We display an evidence of the symmetry breaking 
in the Heisenberg and {\it XY} antiferromagnets. 

First we discuss the necessary conditions for symmetry breaking to occur. 
Bernu {\it et al.}\upcite{Bernu} have also discussed these matters. 
We discuss the conditions of the creation of the spontaneous sublattice 
magnetization and of the spontaneous chirality separately. 

When the symmetry breaking occurs in the thermodynamic
limit, the low-lying states which satisfy the following two conditions
should merge into the ground states\footnote{For a definition of the 
ground state in quantum spin systems, see refs.~\onlinecite{Bratteli} and 
\onlinecite{KomaB}.} in the infinite-volume limit.
\begin{itemize}
\item[(1)] The energy (per site) is the same as that of the ground state 
in the infinite-size limit. 
\item[(2)] The spatial-symmetry is the same as that of the ground state of the 
classical model, i.e. the state has sublattice-translational invariance 
and $C_{3V}$-invariance. 
\end{itemize}
In the thermodynamic limit, these low-lying states form pure ground states. 
In general, all the pure ground states give the same physical
quantities. The sublattice magnetization and the chirality per site should be 
the same throughout the ground states. From this consideration and the 
discussions in Section~3, we obtain 
the necessary condition of the creation of the spontaneous sublattice
magnetization as follows: 
\begin{itemize}
\item[(3)] A number of low-lying states of the type $\alpha$ satisfy the 
condition (1) and they have the same macroscopic value of long-range 
order of the sublattice magnetization. 
\end{itemize}
The necessary condition of the creation of the spontaneous chirality is as
follows: 
\begin{itemize}
\item[(4)] Pairs of low-lying states of the type $\alpha$ and of $\beta$ 
satisfy the condition (1) and they have the same macroscopic value of 
long-range order of the chirality. 
\end{itemize}

Bernu {\it et al}.\upcite{Bernu} studied the low-lying states of the Heisenberg 
model on the triangular lattice. 
They found many low-lying states which satisfy the conditions (1) and (2), and  
found that these states have similar macroscopic values for the sublattice 
magnetization. 
They estimated the sublattice magnetization of the ground state in the 
infinite-volume limit for the Heisenberg model. 
Leung and Runge\upcite{Leung} estimated the sublattice magnetization and 
chirality of the ground state for the {\it XY} and Heisenberg models. 

To examine these conditions explicitly, we study the energy gaps, the
sublattice magnetization and the chirality of the low-lying states in the
$S=1/2$ {\it XY} and Heisenberg models, and estimate the physical values of the 
low-lying states in the infinite-volume limit. 
The systems of the size $N=9$, 12, 21, and 27 with periodic-boundary 
conditions are studied, using the exact-diagonalization method. 
We study the low-lying states which belong to the subspaces of 
$S^z_{\rm total}=0$ (or $0.5$) and of $S^z_{\rm total}=1$ (or $1.5$), and 
which satisfy the condition (2). 
(The ground state and some low-lying excited states really belong to these 
subspaces.) As we have indicated in Section 2, there exist the 
$C_6$-symmetric (type $\alpha$) and $C_6$-antisymmetric (type $\beta$) states. 
In many cases, we studied the lowest state of each subspace.
There are exceptions: In the space of $S^z_{\rm total}=0$ for $N=12$ and
$S^z_{\rm total}=1.5$ for $N=21$ of the Heisenberg model, 
we chose the first-excited state of the type $\beta$, since the first-excited 
state of the type $\alpha$ and the lowest state of the type $\beta$ are 
degenerate and form a paired doublet, and the ground state (the type $\alpha$)
and the first excited state of the type $\beta$ are in pairs. 

The lowest state in the $S^z_{\rm total}=0$ and $C_6$-symmetric (type $\alpha$) 
subspace is the ground state, which has been already calculated up to $N=36$
in refs.~\onlinecite{Bernu} and \onlinecite{Leung}. We used the data for 
$N=36$ in ref.~\onlinecite{Leung}. 

First we discuss the energy gap. Here the energy gap is defined as the 
difference between the total energies of two states. 
The energy gap between the lowest state of the type $\alpha$ and that of 
$\beta$ is shown in Fig.~2. 
In many cases the states of the type $\alpha$ and
$\beta$ are degenerate. The energy gap between the ground state (the type 
$\alpha$) and the lowest state 
(the type $\alpha$) of the $S^z_{\rm total}=1$ subspace is plotted in
Fig.~3. It decreases proportionally to $N^{-1}$. 
As we have discussed in Section~\ref{sec:SB_mechanism}, 
the former energy gap is relevant to the symmetry breaking of the chirality and 
the latter is to the breakdown of the rotational symmetry 
which creates the spontaneous sublattice magnetization. 
The latter energy gap corresponds to the singlet-triplet gap in the Heisenberg
model, which was reported in ref.~\onlinecite{Leung}. 

It was proved that, if there is a long-range order, the energy gap decreases 
in the $N^{-1}$ form or faster than it.\upcite{Horsch,KomaB}
Thus the above calculated results indicate an evidence for the existence of
long-range orders in the {\it XY} and Heisenberg antiferromagnets. 

Next we consider the sublattice magnetization which is observed with the
operator 
\begin{equation}
M=\sum_{i\in A} S_i^x 
+\sum_{i\in B}\Bigl(-{1\over 2} S_i^x +{\sqrt3 \over 2}S_i^y\Bigr)
+\sum_{i\in C}\Bigl(-{1\over 2} S_i^x -{\sqrt3 \over 2}S_i^y\Bigr).
\end{equation}
We calculated the long-range order of the sublattice magnetization 
$\langle M^2 \rangle/N^2$. The values of the
spontaneous sublattice magnetization $m$ are estimated, using the relation
$m=\lim_{N\uparrow\infty} \sqrt{6\langle M^2 \rangle}/N$ for the Heisenberg
model and $m=\lim_{N\uparrow\infty} 2\sqrt{\langle M^2 \rangle}/N$ for the 
{\it XY} model. We show a derivation of these relations in Appendix B. 
We fit the data of the sublattice magnetization $m$ to the $N^{-1/2}$ form. 
This finite-size correction was derived from the spin-wave theory\upcite{Huse}
and using the effective Hamiltonian for a large spin\upcite{Azaria1993}. 
(We also fit the long-range correlation $\langle M^2 \rangle/N^2$ to the 
$N^{-1/2}$ form, using the data for 
$N=3$, 9, 12, 21, and 27, and found that the correction term of $O(N^{-1})$ 
in the fitting of $\langle M^2 \rangle/N^2$ 
is larger than that in the fitting of $m$. 
Therefore we adopted the fitting of $m$.) 
The results for the {\it XY} model are shown in Fig.~4. 
We extrapolated in the $N^{-1/2}$ form, using the data for $N=9$, 12, 21, and 
27. For the ground state we also used the data for $N=36$ which was reported 
in ref.~\onlinecite{Leung}. The values 
in the infinite-volume limit are shown in Table~\ref{Table:XY}. 
They coincide with each other. Thus the condition~(3) is satisfied 
in the {\it XY} model. 
The data for the Heisenberg model are shown in Fig.~5.
The values in the infinite-size limit are shown in Table~\ref{Table:H}.  
They are almost equal. Thus the condition~(3) is also satisfied in the
Heisenberg model. 
All the values are nonvanishing and close to the result by the spin-wave 
theory. However, as it is reported in ref.~\onlinecite{Leung}, if we fit the
data $\langle M^2 \rangle/N^2$ to $N^{-1/2}$, we obtain the estimate 
$m\simeq 0$.
The values seriously depend on the fitting form. 
Thus it is still hard to conclude the existence of the long-range
order in the Heisenberg model. 
\begin{table}[th]
\caption{Estimates of the sublattice magnetization $m$ and the chirality $q$ 
for the $S=1/2$-{\it XY} antiferromagnet on the triangular lattice.}
\label{Table:XY}
\begin{center}
\begin{tabular}{lcc}
\hline
\hline
                                 & $m$   &  $q$  \\
\hline
$S^z_{\rm total}$=0, Type $\alpha$   &  0.41 &  0.76 \\
$S^z_{\rm total}$=0, Type $\beta$    &  0.39 &  0.76 \\
$S^z_{\rm total}$=1, Type $\alpha$   &  0.42 &  0.77 \\
$S^z_{\rm total}$=1, Type $\beta$    &  0.41 &  0.78 \\
\hline
Spin-wave expansion     & 0.437\upcite{Leung} &  0.798\upcite{Momoi} \\
\hline
\hline
\end{tabular}
\end{center}
\end{table}

\begin{table}[hb]
\caption{Estimates of the sublattice magnetization $m$ and the chirality $q$ 
for the $S=1/2$-Heisenberg antiferromagnet on the triangular lattice.}
\label{Table:H}
\begin{center}
\begin{tabular}{lcc}
\hline
\hline
                                 & $m$   &  $q$  \\
\hline
$S^z_{\rm total}=0$, Type $\alpha$   &  0.24 &  0.21 \\
$S^z_{\rm total}=0$, Type $\beta$    &  0.22 &  0.16 \\
$S^z_{\rm total}=1$, Type $\alpha$   &  0.24 &  0.54 \\
$S^z_{\rm total}=1$, Type $\beta$    &  0.38 &  0.73 \\
\hline
Spin-wave expansion       & 0.250\upcite{Miyake}  &  0.405\upcite{Momoi} \\
\hline
\hline
\end{tabular}
\end{center}
\end{table}

Finally we consider the chiral order, which is observed with the operator 
\begin{equation}
\mbox{\boldmath $Q$}={2\over\sqrt3}\sum_{<i\to j>}\mbox{\boldmath $S$}_i\times 
\mbox{\boldmath $S$}_j,
\end{equation}
where the symbol $i\to j$ goes from the sublattice A to B, B to C, and C to A.
We calculated the long-range order of the chirality 
$\langle (Q^z)^2 \rangle/N^2$. The values of the spontaneous chirality $q$ are
estimated using the relations $q=\lim_{N\uparrow\infty} 
\sqrt{3 \langle (Q^z)^2 \rangle}/N$ for the Heisenberg model and 
$q=\lim_{N\uparrow\infty} \sqrt{\langle (Q^z)^2 \rangle}/N$ for the {\it XY}
model. 
We show the results for the {\it XY} model in Fig.~6 and for the 
Heisenberg model in Fig.~7. 
For the {\it XY} model we extrapolated the expectation values of the chirality 
in the $N^{-3/2}$ form, which we derive from the finite-size correction 
of the spin-wave theory\upcite{Momoi}.
On the other hand, in the Heisenberg model the finite-size correction behaves 
in the $N^{-1/2}$ form, as discussed by Azaria {\it et al}.\upcite{Azaria1993}
This difference of the correction terms comes from the fact that in the 
Heisenberg model the chirality 
is sensitive to spin-wave fluctuations of long wave-length, 
while in the {\it XY} model the chiral order is stable against them. 
The extrapolated values are shown in Tables~\ref{Table:XY} 
and~\ref{Table:H}. In the
{\it XY} model the values are consistent with each other and
nonvanishing, which suggests the existence of the chiral order. 
Thus the condition~(4) is satisfied. In 
the Heisenberg model the values do not agree with each other. 
It is necessary to calculate larger systems to conclude that these low-lying
states have the same chirality. We can not fit the data to a form including
higher-order terms, since the size of the data is too small. Thus we can not
still conclude the existence of the long-range order of the chirality in
the Heisenberg model. 

To test the mechanism 
of symmetry-breaking of the chirality which we have discussed in
Section~\ref{sec:SB_mechanism}, we construct the following 
state 
\begin{equation}\label{state:SB_ch}
\vert \phi \rangle={1\over\sqrt2}(\vert \alpha \rangle+ i\vert \beta\rangle), 
\end{equation}
where $\vert\alpha\rangle$ ($\vert\beta\rangle$) denotes the lowest state 
of the type $\alpha$ ($\beta$) of each $S^z_{\rm total}$ subspace. 
We observe the spontaneous chirality of this state. 
The expectation value of the chirality is calculated through the relation  
\begin{equation}
\langle \phi \vert Q^z \vert \phi \rangle 
=i\langle \alpha \vert Q^z \vert\beta\rangle.
\end{equation}
The estimates in the {\it XY} model are shown in Fig.~8. 
They are almost equal to the values which are shown in Fig.~6. 
This coincidence suggests correctness of the mechanism of the 
chiral symmetry breaking which was discussed in Section~\ref{sec:SB_mechanism}. 
In the Heisenberg model we multiply the factor $\sqrt3$ to the calculated
values, since the symmetry is not fully broken in the 
state~(\ref{state:SB_ch}); the state~(\ref{state_appro2H}) should be 
properly used instead of (\ref{state:SB_ch}).  
The estimates are shown in Fig.~9.  
Thus we can break the chiral symmetry by making a linear combination of 
low-lying states of the type $\alpha$ and of the type $\beta$. 

From the above results we summarize that 
the necessary conditions (3) and (4) are satisfied in the {\it XY} model. 
In the Heisenberg model the necessary conditions are almost satisfied, though 
there still remains ambiguity in the point that the low-lying states have the
same properties. 
In any case, in the above results there is no evidence which contradicts 
the occurrence of symmetry breaking.

\setcounter{equation}{0}
\section{Summary }\label{sec:summary}
To summarize, we studied the low-lying states 
of quantum antiferromagnets on a triangular lattice to clarify what kind of
roles they play in the symmetry breaking 
and to show an evidence of occurrence of the symmetry breaking. 

We gave approximate forms of low-lying states, which have two-fold structures. 
It was found that these approximate states resemble the true states in various
properties. We classified the low-lying states by the
eigenvalues of $C_6$, namely $C_6$-symmetric and $C_6$-antisymmetric. 
States of both types exist in pairs in the true
low-lying spectrum. We discussed how rearrangements of the 
low-lying states of two types bring on the symmetry breaking. 
The spontaneous chirality is created by taking a linear combination of 
pairs of low-lying state of the type $\alpha$ and of $\beta$. 
The spontaneous sublattice magnetization is obtained by 
making a linear combination of low-lying states of the same type. 

To display an evidence of the occurrence of symmetry breaking, we studied 
low-lying states of finite systems using the exact-diagonalization method.
We studied whether the necessary conditions of the symmetry breaking are
satisfied or not. We found that the 
conditions are satisfied in the {\it XY} antiferromagnet. The estimate of the 
sublattice magnetization is $m=0.41$ and of chirality is $q=0.77$. These values
are less than the spin-wave results by $7\%$ and $4\%$ respectively. 
In the Heisenberg antiferromagnet the conditions are almost satisfied, 
although the estimates of the sublattice magnetization and the chirality do not 
converge well for lack of the system-size. 
The sublattice magnetization is estimated as $m=0.22\sim 0.38$ and the 
chirality is $q=0.16 \sim 0.73$, where the results from the spin-wave expansion 
are $m=0.25$ and $q=0.45$. Since the extrapolated results seriously 
depend on fitting forms, it is still hard in the present study to conclude
definitely the existence of the long-range order in the Heisenberg model. 

\setcounter{equation}{0}
\appendices
\section*{Appendix A. \ \ Low-lying states of the ferromagnetic XXZ model}
We show the uniqueness of the ground state and give spatial quantum numbers of
the low-lying states of the spin $S$ $XXZ$ ferromagnet on finite systems. 
We discuss the model 
on the finite-volume triangular lattice $\Lambda$ with a periodic 
boundary condition. 
For simplicity, we restrict the number of sites to an even integer. 
The Hamiltonian is given by 
\begin{equation}\label{Hamiltonian_ferro}
H_F=-J\sum_{\langle i,j \rangle \in \Lambda}(S_i^x S_j^x+S_i^y S_j^y 
+ \eta S_i^z S_j^z),
\end{equation}
where $J>0$ and $\eta<1$. The summation runs over all the nearest-neighbor 
sites. 

The Hamiltonian (\ref{Hamiltonian_ferro}) has nonpositive off-diagonal
elements. The basis which have the same $S^z_{\rm total}$ are connected by 
the elements
of the Hamiltonian. From the Perron-Frobenius theorem, 
the lowest eigenstate in each $S^z_{\rm total}$ subspace is unique 
and it has nonnegative coefficients. 

Here we show that the ground state exists uniquely in the $S^z_{\rm total}=0$ 
subspace. We consider the following Hamiltonian 
\begin{equation}\label{Hamiltonian_general}
H_F^\prime=-\sum_{ i,j \in \Lambda}J_{ij}
(S_i^x S_j^x+S_i^y S_j^y + \eta S_i^z S_j^z),
\end{equation}
where $J_{ij} \ge 0$ for all $i$ and $j$. Affleck and Lieb\upcite{Affleck} 
showed that, in the
antiferromagnetic $XXZ$ chain, the ground state exists uniquely in the
$S^z_{\rm total}=0$
subspace. As they did, we transform the Hamiltonian (\ref{Hamiltonian_general}) 
using the unitary operator 
\begin{equation}
W=\exp\Bigl(i{\pi \over 2} \sum_i S_i^x \Bigr)
\end{equation}
into the form
\begin{equation}
W^\dagger H_F^\prime W=-J\sum_{ i,j \in \Lambda}
\Bigl\{ {1\over 4}(1+\eta)(S_i^+ S_j^- + S_i^- S_j^+) 
+{1\over 4}(1-\eta)(S_i^+ S_j^+ + S_i^- S_j^-) + S_i^z S_j^z \Bigr\}.
\end{equation}
In the region $-1<\eta<1$, this transformed 
Hamiltonian has nonpositive off-diagonal elements. It has two connected
blocks, i.e. all basis in the subspaces with even $S^z_{\rm total}$ are 
connected by
the elements of the Hamiltonian and those with odd $S^z_{\rm total}$ 
are connected. 
From the Perron-Frobenius theorem, the lowest eigenstate in
each connected subspace is unique. Thus we find that the number of the 
ground states 
is at most two in the original model (\ref{Hamiltonian_general}). 
The lowest states in the $S^z_{\rm total}=n$ and $S^z_{\rm total}=-n$ subspaces 
are degenerate. When level crossing of the ground state
occurs, more than two ground states must exist, which contradicts the
above result. Thus in the whole region of various parameters 
$\{$~$J_{ij}$~$\}$, the ground state
should exist in the same $S^z_{\rm total}$ subspace. 

For the region $\eta<-1$, as Affleck and Lieb did, the above results can be
shown using the unitary operator 
\begin{equation}
W=\exp\Bigl(i{\pi \over 2} \sum_i S_i^y \Bigr).
\end{equation}

To obtain the eigenvalue $S^z_{\rm total}$ of the ground state, we
consider the case $J_{ij}=J$ for all $i$ and $j$. The Hamiltonian
can be written in the form 
\begin{equation}
H_F^\prime=-J\{(\mbox{\boldmath $S$})^2 - (1-\eta) (S^z)^2 \},
\end{equation}
where \mbox{\boldmath $S$}$=(S^x,S^y,S^z)$ and 
$S^\alpha=\sum_{i\in \Lambda}S^\alpha_i$ $(\alpha=x,y,z)$. 
The ground state of this model is unique and it has the 
eigenvalues $S_{\rm total}=SN$ and $S^z_{\rm total}=0$.\upcite{Suzuki}
From the above results, we find that the ground state of the original model
(\ref{Hamiltonian_ferro}) in the regions $\eta <-1$ and $-1<\eta<1$ exists 
uniquely in the $S^z_{\rm total}=0$ subspace. 

This argument can be extended to the ferromagnetic $XXZ$ model on any lattice.
The ferromagnet on a bipartite lattice is equivalent to the $XXZ$
antiferromagnet with the $z$-component anisotropy $-\eta$. In this case the
above results are consistent with what were proved by Lieb and 
Mattis\upcite{Lieb}, and Affleck and Lieb\upcite{Affleck}. 

Lastly, we give the spatial quantum numbers of the lowest state in each
$S^z_{\rm total}$ subspace. The Hamiltonian (\ref{Hamiltonian_ferro}) is 
invariant under any translation, rotation, and reflection. 
As we have shown in the above, the
lowest state is nondegenerate and it has nonnegative coefficients. From this
fact, we find that the lowest state has the eigenvalue $1$ for any translation,
rotation, and reflection, i.e. it is translationally invariant and
$C_{6V}$-invariant. 

\setcounter{equation}{0}
\setcounter{section}{2}
\section*{Appendix B. \ \ Relations between long-range order and symmetry
breaking }
We discuss the relation between spontaneous 
symmetry breaking $m$ and the long-range order parameter $\sigma$. 
It has been discussed\upcite{Miyashita1989,KomaA} that the relation is given by 
$m=\sqrt3 \sigma$ for the Heisenberg
antiferromagnet and by $m=\sqrt2 \sigma$ for the $XY$ antiferromagnet on 
bipartite lattices. 
But these relations are not valid in the antiferromagnets on the triangular 
lattice. 
Here we show that, in the antiferromagnets on the triangular lattice, 
the factor becomes $\sqrt2$ times as large as the above. 

Our arguments are based on the assumption that the symmetric infinite-volume
ground state, (\ref{def:mixed_state}), is decomposed into the pure ground 
states. Koma and Tasaki\upcite{KomaA} used this decomposition to explain 
the factor $\sqrt3$ for the Heisenberg antiferromagnets on bipartite systems. 

The spontaneous sublattice magnetization $m$ is defined as 
\begin{equation}
m\equiv \lim_{B\downarrow 0}\lim_{N\uparrow \infty}\lim_{\beta\uparrow \infty}
{1\over N}\frac{\mbox{Tr}[M \exp\{-\beta (H-BM)\}]}
{\mbox{Tr}[\exp\{-\beta (H-BM)\}]}.
\end{equation}
Using the sublattice-translational invariance and equivalence of the sublattice
magnetization, we have 
\begin{equation}
m=\tilde{\omega} (S^x_0), 
\end{equation}
where the state $\tilde{\omega} (\cdots)$ is defined by (\ref{def:pure_state}). 
The long-range order parameter of the sublattice 
magnetization $\sigma_m$ is defined by 
\begin{equation}
{\sigma_m}^2 \equiv \lim_{N\uparrow \infty}\lim_{\beta\uparrow \infty}
{1\over N^2}\frac{\mbox{Tr}[M^2 \exp(-\beta H)]}
{\mbox{Tr}[\exp(-\beta H)]}.
\end{equation}
Using the rotational and translational invariance, we obtain 
\begin{eqnarray}\label{def:LRO_m}
{\sigma_m}^2 = \lim_{\vert r \vert \uparrow \infty}
{1\over 3}&\{& \omega (S^x_0 S^x_r)
+\omega (-{1\over2}S^x_0 S^x_{r+\mbox{\boldmath \scriptsize $e$}_1}
         +{\sqrt3\over2}S^x_0 S^y_{r+\mbox{\boldmath \scriptsize $e$}_1})
\nonumber\\
& &+\omega (-{1\over2}S^x_0 S^x_{r+\mbox{\boldmath \scriptsize $e$}_2}
         -{\sqrt3\over2}S^x_0 S^y_{r+\mbox{\boldmath \scriptsize $e$}_2})\},
\end{eqnarray}
where the state $\omega$ is defined by (\ref{def:mixed_state}). The sites 0 and 
$r$ belong to the sublattice A and {\boldmath $e_1$} ({\boldmath $e_2$})
denotes the unit lattice vector, which belongs to the sublattice B (C). 

We also define the spontaneous chirality $q$ as 
\begin{equation}
q\equiv \lim_{B\downarrow 0}\lim_{N\uparrow \infty}\lim_{\beta\uparrow \infty}
{1\over N}\frac{\mbox{Tr}[Q^z \exp\{-\beta (H-BM)\}]}
{\mbox{Tr}[\exp\{-\beta (H-BM)\}]}
=\tilde{\omega} (Q^z(0))
\end{equation}
and the long-range order parameter of the chirality $\sigma_q$ as 
\begin{equation}\label{def:LRO_ch}
{\sigma_q}^2 \equiv \lim_{N\uparrow \infty}\lim_{\beta\uparrow \infty}
{1\over N^2}\frac{\mbox{Tr}[(Q^z)^2 \exp(-\beta H)]}
{\mbox{Tr}[\exp(-\beta H)]}
=\lim_{\vert r \vert\uparrow \infty}\omega (Q^z(0)Q^z(r)),
\end{equation}
where the operator $Q^z(r)$ is the {\it z}-component of the chiral 
order-parameter operator on the unit triangular cell, 
\begin{equation}
\mbox{\boldmath $Q$}(r)={2\over \sqrt3}
(\mbox{\boldmath $S$}_r\times
\mbox{\boldmath $S$}_{r+\mbox{\boldmath \scriptsize $e$}_1}
+\mbox{\boldmath $S$}_{r+\mbox{\boldmath \scriptsize $e$}_1}\times
\mbox{\boldmath $S$}_{r+\mbox{\boldmath \scriptsize $e$}_2}
+\mbox{\boldmath $S$}_{r+\mbox{\boldmath \scriptsize $e$}_2}\times
\mbox{\boldmath $S$}_r ).
\end{equation}

From now we derive the relations between $m$ and $\sigma_m$, and between $q$
and $\sigma_q$.
Our results are as follows: 
For the {\it XY} model 
\begin{equation}\label{relation:SB_LRO_XY}
m=2\sigma_m \mbox{\hspace{1cm} and \hspace{1cm}} q=\sigma_q ,
\end{equation}
and for the Heisenberg model 
\begin{equation}\label{relation:SB_LRO_H}
m=\sqrt6\sigma_m \mbox{\hspace{1cm} and \hspace{1cm}} q=\sqrt3\sigma_q.
\end{equation}
The relation $m=\sqrt6\sigma_m$ for the Heisenberg model has been already 
used by Bernu {\it et al.}\upcite{Bernu}. 

To derive these relations, we use the following standard arguments. 
It is widely believed that the mixed state $\omega$ can be naturally 
decomposed into pure equilibrium states,\upcite{Bratteli} 
\begin{equation}\label{eq:decomposition}
\omega (\cdots)=\int d\alpha \omega_\alpha (\cdots),
\end{equation}
where $\{\omega_\alpha\}$ denote the pure states and the parameter $\alpha$
describes the properties of them. 
The state $\tilde{\omega}$, which is defined in eq.~(\ref{def:pure_state}), 
is one of $\omega_\alpha$. 
The pure states have the cluster property,\upcite{Bratteli} 
\begin{equation}\label{eq_cluster}
\omega_\alpha (A_0 B_r) 
\mathop{\hbox to 1.5cm{\rightarrowfill}}_{\vert r \vert\uparrow \infty} 
\omega_\alpha (A_0)\omega_\alpha (B_r)
\end{equation}
for any local operators $A$ and $B$. 

\subsection{XY-like model}
In the {\it XXZ} model with $-0.5<\Delta <1$, it is expected that the
decomposition~(\ref{eq:decomposition}) is of the form 
\begin{equation}\label{eq_decompXY}
\omega (\cdots)={1 \over 4\pi}\int_0^{2\pi} d\theta 
\{ \omega_{\theta,+} (\cdots) + \omega_{\theta,-} (\cdots)\},
\end{equation}
where $\omega_{\theta,+}$($\omega_{\theta,-}$) denote the
states which have positive(negative) chirality and in which vectors of spins on 
the sublattice A form the angle $\theta$ with the $x$-axis. 
The state $\tilde{\omega}$ corresponds to 
$\omega_{\theta=0,+}$. 
The pure states have following expectation values of the sublattice 
magnetization 
\begin{eqnarray}
\omega_{\theta,+} \Bigl( S_0^x \Bigr)=m\cos \theta,
& &\omega_{\theta,-} \Bigl( S_0^x \Bigr)=m\cos \theta,\\
\omega_{\theta,+} \Bigl( 
-{1\over2}S_{\mbox{\boldmath \scriptsize $e_1$}}^x
+{\sqrt3 \over2}S_{\mbox{\boldmath \scriptsize $e_1$}}^y
\Bigr)=m\cos \theta,
& &\omega_{\theta,-} \Bigl( 
-{1\over2}S_{\mbox{\boldmath \scriptsize $e_1$}}^x
+{\sqrt3 \over2}S_{\mbox{\boldmath \scriptsize $e_1$}}^y
\Bigr)=m\cos (\theta-{2\pi\over3}), \nonumber\\
\omega_{\theta,+} \Bigl( 
-{1\over2}S_{\mbox{\boldmath \scriptsize $e_2$}}^x
-{\sqrt3 \over2}S_{\mbox{\boldmath \scriptsize $e_2$}}^y
\Bigr)=m\cos \theta,
& &\omega_{\theta,-} \Bigl( 
-{1\over2}S_{\mbox{\boldmath \scriptsize $e_2$}}^x
-{\sqrt3 \over2}S_{\mbox{\boldmath \scriptsize $e_2$}}^y
\Bigr)=m\cos (\theta+{2\pi\over3}),\nonumber
\end{eqnarray}
and the chirality 
\begin{equation}
\omega_{\theta,\pm}(Q^z(0))=\pm q. 
\end{equation}

Using the decomposition~(\ref{eq_decompXY}) and the 
property~(\ref{eq_cluster}), the long range order
parameter of the sublattice magnetization, (\ref{def:LRO_m}), is transformed as 
\begin{eqnarray}
{\sigma_m}^2 
&=&\lim_{\vert r \vert \uparrow \infty}{1 \over 12\pi}\int_0^{2\pi} d\theta
\Bigl\{\omega_{\theta,+} (S^x_0 S^x_r)
+\omega_{\theta,+} (-{1\over2}S^x_0 S^x_{r+\mbox{\boldmath \scriptsize $e$}_1}
         +{\sqrt3\over2}S^x_0 S^y_{r+\mbox{\boldmath \scriptsize $e$}_1})
\nonumber\\
& & \mbox{  } 
+\omega_{\theta,+} (-{1\over2}S^x_0 S^x_{r+\mbox{\boldmath \scriptsize $e$}_2}
         -{\sqrt3\over2}S^x_0 S^y_{r+\mbox{\boldmath \scriptsize $e$}_2})
+\omega_{\theta,-} (S^x_0 S^x_r)
\nonumber\\
& & \mbox{  }
+\omega_{\theta,-} (-{1\over2}S^x_0 S^x_{r+\mbox{\boldmath \scriptsize $e$}_1}
         +{\sqrt3\over2}S^x_0 S^y_{r+\mbox{\boldmath \scriptsize $e$}_1})
+\omega_{\theta,-} (-{1\over2}S^x_0 S^x_{r+\mbox{\boldmath \scriptsize $e$}_2}
         -{\sqrt3\over2}S^x_0 S^y_{r+\mbox{\boldmath \scriptsize $e$}_2})
\Bigr\}\nonumber\\
&=&{1 \over 12\pi}\int_0^{2\pi} d\theta \nonumber\\
& &\Bigl[\omega_{\theta,+} (S^x_0) \Bigl\{ \omega_{\theta,+} (S^x_0)
+\omega_{\theta,+} (-{1\over2}S^x_{\mbox{\boldmath \scriptsize $e$}_1}
                    +{\sqrt3\over2}S^y_{\mbox{\boldmath \scriptsize $e$}_1})
+\omega_{\theta,+} (-{1\over2}S^x_{\mbox{\boldmath \scriptsize $e$}_2}
             -{\sqrt3\over2}S^y_{\mbox{\boldmath \scriptsize $e$}_2})\Bigr\}
\nonumber\\
& &+\omega_{\theta,-} (S^x_0) \Bigl\{ \omega_{\theta,-} (S^x_0)
+\omega_{\theta,-} (-{1\over2}S^x_{\mbox{\boldmath \scriptsize $e$}_1}
                    +{\sqrt3\over2}S^y_{\mbox{\boldmath \scriptsize $e$}_1})
+\omega_{\theta,-} (-{1\over2}S^x_{\mbox{\boldmath \scriptsize $e$}_2}
                    -{\sqrt3\over2}S^y_{\mbox{\boldmath \scriptsize $e$}_2})
\Bigr\}\Bigr]\nonumber\\
&=& {m^2 \over4\pi}\int^{2\pi}_0 d\theta \cos^2 \theta
={m^2\over4}.
\end{eqnarray}
Thus we obtain the relation $m=2 \sigma_m$.

In the same way, 
the long-range order parameter of the chirality, (\ref{def:LRO_ch}), is 
estimated as 
\begin{eqnarray}
{\sigma_q}^2
&=&\lim_{\vert r \vert\uparrow \infty}{1 \over 4\pi}\int_0^{2\pi} d\theta 
\Bigl\{\omega_{\theta,+} (Q^z(0)Q^z(r)) +\omega_{\theta,-} (Q^z(0)Q^z(r))
\Bigr\}\nonumber\\
&=&{1 \over 4\pi}\int_0^{2\pi} d\theta 
\{\omega_{\theta,+} (Q^z(0))^2 +\omega_{\theta,-} (Q^z(0))^2 \}
= q^2.
\end{eqnarray}
Thus we get the relation $q=\sigma_q$. 

\subsection{Heisenberg model}
For the Heisenberg model it is expected that the mixed state 
$\omega (\cdots)$ is decomposed as
\begin{equation}\label{eq_decompH}
\omega (\cdots)={1\over 8\pi^2}\int d\Omega 
\int_0^{2\pi} d\phi 
\omega_{\mbox{\boldmath \scriptsize $\Omega$},\phi} (\cdots),
\end{equation}
where the vector $\mbox{\boldmath $\Omega$}$ is perpendicular to the plane 
to which all spins are
parallel, and the angle $\phi$ denotes rotation 
with respect to the vector $\mbox{\boldmath $\Omega$}$. 
The direction of the vector
$\mbox{\boldmath $\Omega$}$ and that of the chirality are the same, 
which are described with the spherical coordinates ($\theta,\varphi$). 
The state $\tilde{\omega}$ corresponds to the state 
$\omega_{\varphi=0 \atop \phi=0}$. 
The pure states have following expectation values of the sublattice 
magnetization
\begin{eqnarray}
& &\omega_{\mbox{\boldmath \scriptsize $\Omega$},\phi} 
\Bigl(S_0^x \Bigr) 
=m(\cos\varphi \cos\theta \cos\phi - \sin\varphi \sin\phi),\\
& &\omega_{\mbox{\boldmath \scriptsize $\Omega$},\phi} 
\Bigl(-{1\over2}S_{\mbox{\boldmath \scriptsize $e_1$}}^x
+{\sqrt3\over2}S^y_{\mbox{\boldmath \scriptsize $e_1$}} \Bigr) \nonumber\\
& & \mbox{  } =m\Bigl\{
\cos\Bigl(\varphi-{2\pi\over3}\Bigr) \cos\theta 
\cos\Bigl(\phi+{2\pi\over3}\Bigr) 
- \sin\Bigl(\varphi-{2\pi\over3}\Bigr) 
\sin\Bigl(\phi+{2\pi\over3}\Bigr)\Bigr\},\nonumber\\
& & \omega_{\mbox{\boldmath \scriptsize $\Omega$},\phi} 
\Bigl(-{1\over2}S_{\mbox{\boldmath \scriptsize $e_2$}}^x
-{\sqrt3\over2}S^y_{\mbox{\boldmath \scriptsize $e_2$}} \Bigr) \nonumber\\
& & \mbox{  }=m\Bigl\{
\cos\Bigl(\varphi+{2\pi\over3}\Bigr) \cos\theta 
\cos\Bigl(\phi-{2\pi\over3}\Bigr) 
- \sin\Bigl(\varphi+{2\pi\over3}\Bigr) 
\sin\Bigl(\phi-{2\pi\over3}\Bigr)\Bigr\},\nonumber
\end{eqnarray}
and the chirality 
\begin{equation}
\omega_{\mbox{\boldmath \scriptsize $\Omega$},\phi}
(\mbox{\boldmath $Q$}(0))
=q\mbox{\boldmath $\Omega$}.
\end{equation}

Using the decomposition~(\ref{eq_decompH}) and the property~(\ref{eq_cluster}), 
the long-range order parameter of the sublattice magnetization is transformed
as 
\begin{eqnarray}
{\sigma_m}^2 
&=&{1\over 24\pi^2}\int d\Omega \int_0^{2\pi} d\phi 
\omega_{\mbox{\boldmath \scriptsize $\Omega$},\phi} (S^x_0) 
\nonumber\\
& & \mbox{  }\times\{\omega_{\mbox{\boldmath \scriptsize $\Omega$},\phi} (S^x_0)
 +\omega_{\mbox{\boldmath \scriptsize $\Omega$},\phi} 
  (-{1\over2}S^x_{\mbox{\boldmath \scriptsize $e$}_1}
   +{\sqrt3\over2}S^y_{\mbox{\boldmath \scriptsize $e$}_1})
 +\omega_{\mbox{\boldmath \scriptsize $\Omega$},\phi} 
  (-{1\over2}S^x_{\mbox{\boldmath \scriptsize $e$}_2}
   -{\sqrt3\over2}S^y_{\mbox{\boldmath \scriptsize $e$}_2})\}
\nonumber\\
&=&{m^2\over 6}.
\end{eqnarray}
Thus we obtain the relation $m=\sqrt6 \sigma_m$. 

Using the decomposition~(\ref{eq_decompH}) and the property~(\ref{eq_cluster}), 
the long-range order parameter of the chirality is transformed as 
\begin{eqnarray}
{\sigma_q}^2 &=& {1\over 3}\lim_{\vert r \vert \uparrow \infty}
\omega (\mbox{\boldmath $Q$} (0) \mbox{\boldmath $Q$} (r)) \\ 
&=&{q^2 \over 24\pi^2}\int d\Omega\int_0^{2\pi}d\phi
 \mbox{\boldmath $\Omega$}^2 
= {q^2 \over3},\nonumber
\end{eqnarray}
where the isotropy of the state is used.
Thus we get the relation $q=\sqrt3 \sigma_q$. 

\section*{Acknowledgments }
The author would like to thank Prof.~M.~Suzuki and Dr.~N.~Hatano for helpful
discussions and critically reading this manuscript. 
The author is also grateful to Dr.~T.~Koma, Prof.~H.~Nishimori and 
Dr.~Y.~Nonomura for fruitful comments. 
Subroutines of TITPACK Ver.2 by Prof.~H.~Nishimori were partly used. 
The numerical calculations have been performed with the HITAC 
S-820/80 of the Computer Center, University of Tokyo and the VAX 6440 of 
the Meson Science Laboratory, Faculty of Science, University of Tokyo.
This study is partially financed by Grant-in-Aid for Scientific Research on
Priority Areas ``Computational Physics as a New Frontier in Condensed Matter
Research'', from the Ministry of Education, Science and Culture, Japan.

\section*{Figure Captions}
\noindent
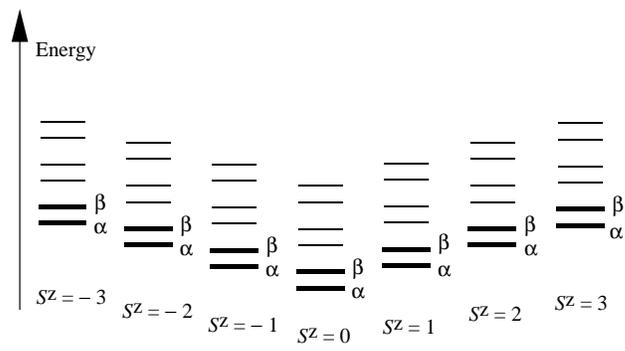
{\bf Figure 1:} Two-fold structures in the spectrum of low-lying states.\\
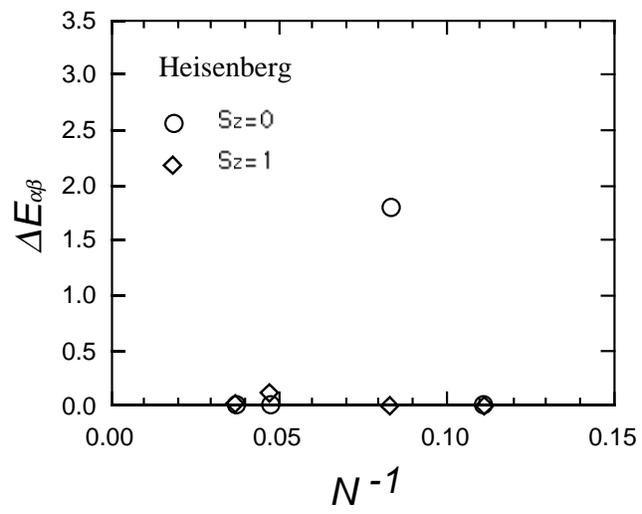
{\bf Figure 2:} Size dependence of the energy gap between the lowest state 
(the type $\alpha$) and the lowest excited state (the type $\beta$) in the 
same $S^z_{\rm total}$-subspace for the Heisenberg model.\\
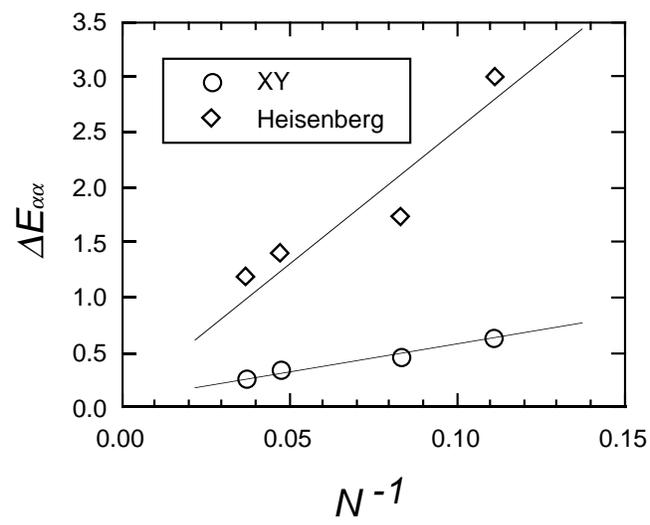
{\bf Figure 3:} Size dependence of the energy gap between the ground state 
(the type 
$\alpha$) and the lowest state (the type $\alpha$) in $S^z_{\rm total}$=1.\\
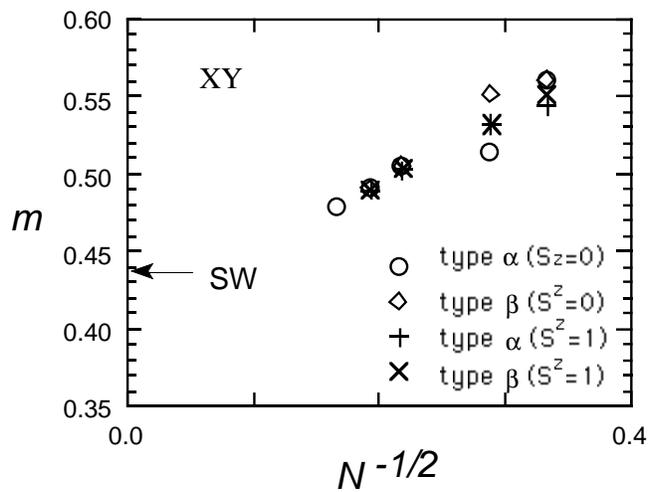
{\bf Figure 4:} Size dependence of the sublattice magnetization of the lowest 
state in each subspace of the {\it XY} model, which is estimated through 
2$\protect\sqrt{\langle M^2 \rangle}/N$. 
The data for $N$=36 is also listed\protect\upcite{Bernu}. 
SW denotes the result from the spin-wave expansion\protect\upcite{Momoi}.\\
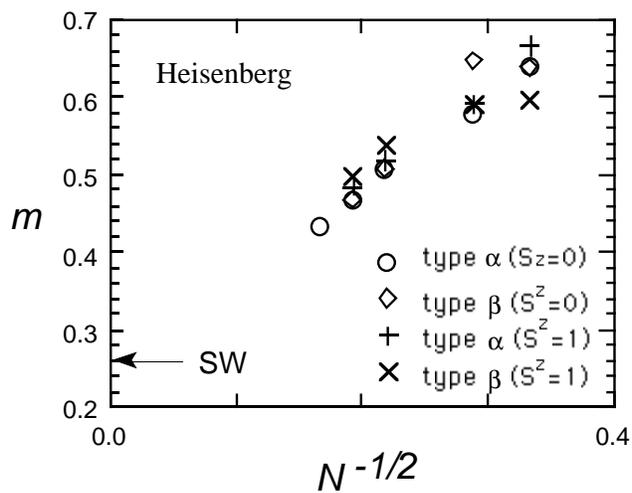
{\bf Figure 5:} Size dependence of the sublattice magnetization of the lowest 
state in each subspace of the Heisenberg model, which is estimated through 
$\protect\sqrt{6\langle M^2 \rangle}/N$. 
The data for {\it N}=36 is also listed\protect\upcite{Bernu}. 
SW denotes the result from the spin-wave expansion\protect\upcite{Miyake}.\\
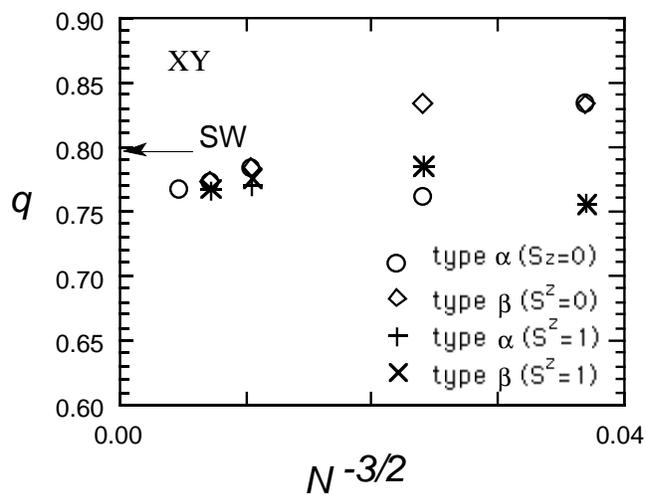
{\bf Figure 6:} Size dependence of the chirality of the lowest state in each 
subspace of the {\it XY} model which is estimated through 
$\protect\sqrt{\langle (Q^z)^2 \rangle}/N$. 
The data for $N=36$ is also listed\protect\upcite{Bernu}. 
SW denotes the result from the spin-wave expansion\protect\upcite{Momoi}.\\
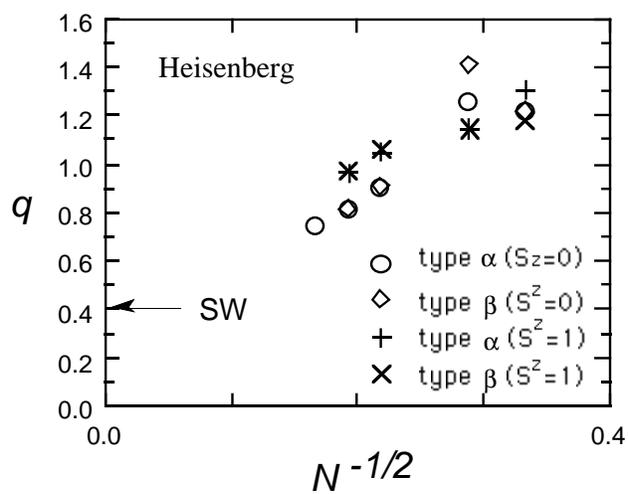
{\bf Figure 7:} Size dependence of the chirality of the lowest state in each 
subspace of the Heisenberg model which is estimated through 
$\protect\sqrt{3\langle (Q^z)^2 \rangle}/N$. 
The data for {\it N}=36 is also listed\protect\upcite{Bernu}. 
SW denotes the result from the spin-wave expansion\protect\upcite{Momoi}.\\
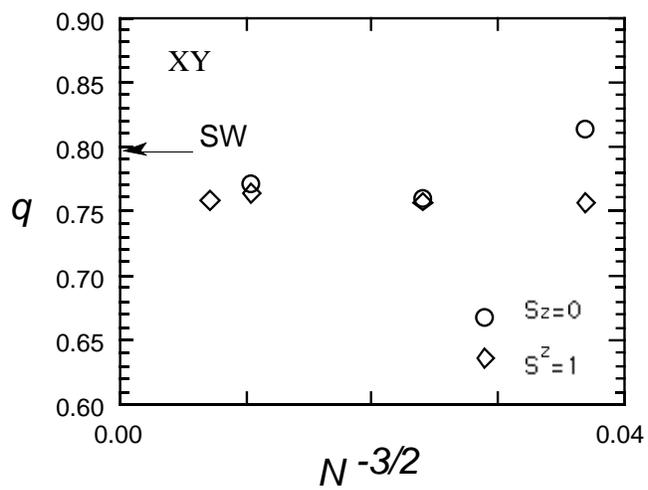
{\bf Figure 8:} Spontaneous symmetry breaking of the chirality of the {\it XY} 
model which is created by a linear combination 
${1\over\protect\sqrt2}(|\alpha\rangle +i|\beta\rangle)$ of the 
low-lying state of the type $\alpha$ and that of $\beta$.\\
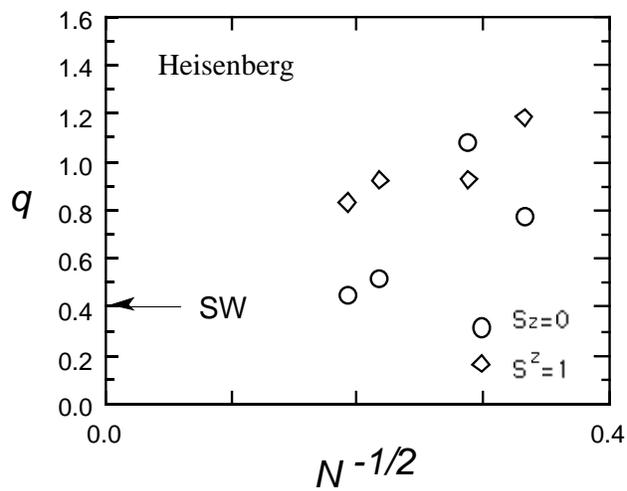
{\bf Figure 9:} Spontaneous symmetry breaking of the chirality of the 
Heisenberg model which is created by a linear combination 
${1\over\protect\sqrt2}(|\alpha\rangle +i|\beta\rangle)$ 
of the low-lying state of the type $\alpha$ and that of $\beta$. 
The values are multiplied by the factor $\protect\sqrt3$.
\end{document}